# TEXT CLASSIFICATION FOR AUTHORSHIP ATTRIBUTION ANALYSIS

M. Sudheep Elayidom[1] , Chinchu Jose[2] , Anitta Puthussery[3] , Neenu K Sasi[4]

[1]Division of Computer Science and Engineering, School of Engineering, Cochin University, Kalamassery, India
[2,3,4]Adi Shankara Institute of Engineering and Technology, Kalady, India

*ABSTRACT*

*Authorship attribution mainly deals with undecided authorship of literary texts. Authorship attribution is useful in resolving issues like uncertain authorship, recognize authorship of unknown texts, spot plagiarism so on. Statistical methods can be used to set apart the approach of an author numerically. The basic methodologies that are made use in computational stylometry are word length, sentence length, vocabulary affluence, frequencies etc. Each author has an inborn style of writing, which is particular to himself. Statistical quantitative techniques can be used to differentiate the approach of an author in a numerical way. The problem can be broken down into three sub problems as author identification, author characterization and similarity detection. The steps involved are pre-processing, extracting features, classification and author identification. For this different classifiers can be used. Here fuzzy learning classifier and SVM are used. After author identification the SVM was found to have more accuracy than Fuzzy classifier. Later combined the classifiers to obtain a better accuracy when compared to individual SVM and fuzzy classifier.*

## 1. INTRODUCTION

Authorship attribution is the process of determining the likely author of a given text document. Applications of authorship attribution include plagiarism detection, resolving disputed authorship etc. Authorship attribution is the technique of determining the author of a text when it is ambiguous who wrote it. It is of use when two or more people argue to have written something or in cases where no one is able to declare that who wrote that document. The complexity of the text authorship problem is obviously exponentially higher the larger number of likely authors. The availability of author text samples is also a major constriction when advancing with this problem. Text authorship attribution engage the following three problems:

1. *The one out of many problem* – identifying the author of a text author from a group of probable or expected authors where the author is always in the group of suspects.
2. *None or one out of many problem* – identifying the author of a text author from a group of probable or expected authors where the author may not be in the group of suspects.

3. *The sole author problem* – estimating the possibility of a given text having been written by the given author or not.

In this work the main focus is on identifying the author of a given text using various steps. Firstly data pre- processing, secondly feature extraction, thirdly classification and finally author identification. Data pre- processing involves text tokenizing and text stemming. Feature extraction involves the process of extracting various features such as top k frequent words, number of punctuations, number of symbols, character count, sentence count, word count and ratio of character count and sentence count. Classification involves the use of three classifier such as fuzzy, SVM and a combination of these two classifiers. Later performance analysis is done to determine the accuracy of each classifiers.

## 2. LITERATURE REVIEW

### 2.1. Creating author fuzzy fingerprints for authorship identification

In this work[1] a fingerprint is extracted from a set of texts. Then that fingerprint is used to identify the author of a distinct text document. This fingerprint is not similar to the biological fingerprint. In the field of computer sciences, fingerprints are usually used to avoid the comparison and transmission of massive data. For example, in order to efficiently verify if a remote file has been altered, a web browser or proxy server can simply get its fingerprint and compare it with the fingerprint of the previously fetched copy. Fingerprints are a swift and compact way to recognize items. To serve the author identification purposes, a fingerprint must be able to capture the identity of that author. In other words, the probability of a conflict, that is, two authors having the same fingerprint, must be small. The fingerprint has also to be robust such that a text should be identified even if the author changes some aspects of the writing style. The idea of identifying text authorship based on an author fingerprint is a very interesting.

When the word frequencies are used as a proxy for the individual behind a particular text, one can collect information on the author and identify other texts. The use of word frequencies is a well-known method. The first step in this method is to gather the top-k word frequencies in all known texts of each known author. An approximated algorithm is used for this purpose since classical exact top-k algorithms are inefficient and require the full list of distinct elements to be kept. The Filtered Space-Saving algorithm is used for this purpose since it provides a fast and compact answer to the top-k problem although it only gives an approximate solution.

After determining the top-k word frequencies, the fingerprint is created by applying a fuzzifying function to the word frequencies. The method of fuzzifying the set of features is based on their order on the top-k list instead of their frequency value. Finally the same calculations are performed for the text being identified and then to compare this text fuzzy fingerprint with all the available author fuzzy fingerprints. The most related fingerprint is chosen and the text is assigned to the author.

### 2.2. Computing frequent and top-k elements present in data streams in an efficient manner

An approximate integrated approach is used for solving two problems. That is finding the most common k elements, and determining the frequent elements in a data stream [2]. It is space efficient and reports both frequent and top-k elements. The top k algorithm returns k elements that have roughly the highest frequencies; and it uses limited space for calculating frequent elements.

For this purpose a counter-based Space- Saving algorithm and its associated Stream-Summary data structure are used. The underlying idea is to maintain partial information of interest. That is only the required m elements will be monitored. The updating of the counters are done in such a manner that, it accurately calculates the frequencies of the major elements. A lightweight data structure is used to keep the elements which are sorted according to their approximate frequencies.

## 2.3. Role of statistical text analysis in authorship attribution

In statistical analysis of literary texts, an objective methodology is applied to the works that have received impressionistic treatment for a extensive time. While in the subjective type of analysis of style of the literature, the literary style of the text is used. But the literary style of the text is not quantifiable, which is an important norm for the judgment. The subjective analysis approach be able to rarely guide to a unique solution which the scholars can accept. The objective components for judgments are provided by the Statistical quantitative methods [3].

In the quantitative analysis approach, the style of an author is characterize numerically by carefully analyzing the style of the text. Then the sets of features in a text that most accurately describe the author's style is determined. Authorship attribution is one of the main applications of stylometry. The stylometry can be defined as the science of measuring literary style. According to many previous studies it is assumed that each author has an inborn style of writing. That writing style is peculiar to that specific author. A established literary scholar can capture the peculiarities in the style of an author by impression. The statisticians recommend to this field is to help quantify the style, and hence to change a subjective method into an objective technique which is called to as "Non-Traditional Stylometry".

The characteristic style of an author can be determined by using certain features. These features may include word length, richness of vocabulary, sentence length, function words so on. Thomas Corwin Mendenhall was the first who undertook extensive work to show that some simple statistical methods may prove useful to solve questions of disputed authorship. He suggested they may also be utilized in comparative language studies, in tracing the growth of a language, in studying the growth of the vocabulary from childhood to manhood, and in other directions. Mendenhall proposed forming relative frequency curve of number of letters per word (word-length), which he called "word-spectrum" or "characteristic curve" as a method of analysis leading to identification or discrimination of authorship. He constructed word-spectra for works of two contemporary novelists called Charles Dickens and the other one is Thackeray, and a few other writers, to show that texts with the same average word-length might possess different spectra. He assumed that every writer makes use of a vocabulary which is specific to himself and the character of which is persistent over time.

## 2.4. Use of Authorship analysis in cybercrime investigation

Authorship analysis is the process of examining the characteristics of a piece of work in order to draw conclusions on its authorship. More specifically, the problem can be broken down into three sub-fields [4]:

• Author Identification find out the likelihood of a particular author having written a piece of work by examining other works produced by that same author.
• Author Characterization summarizes the characteristics of an author and generates the author profile based on that person' s work. Some of these characteristics are gender, culture, educational background, and language familiarity.
• In the Similarity Detection process an author's several pieces of work are compared and find outs whether or not they are produced by a single author without actually identifying the author.

According to this mechanism, three types of features of the message, as well as style markers, content-specific features and, structural features, are extracted. Then after the extraction inductive learning algorithms are used to build feature-based models to identify authorship of illegal messages. In the first step the feature extractor runs on those documents and generates a set of style features, which will be used as the input to/for the learning engine. A feature-based model is then created as the outcome of the learning engine. This model can identify whether a newly found illegal document is written by that suspicious criminal under different usernames or names. Three learning algorithms or classifiers were used for comparison purposes. The first classifier is back propagation neural networks, second one is a decision tree and finally a support vector machine classifier were used.

### 2.5. The Combination of Text Classifiers

A more accurate classification procedure can be developed by combining the outputs of the involved classifiers[5]. Earlier studies of classifier combination have been encouraged mostly by the instinct that superimposed classifiers, that work in related but qualitatively different ways, could leverage the distinct strengths of each method. Classifiers can be combined in different ways. In one method, a text classifier is created from multiple distinct classifiers by selecting the best classifier to use in different situations or contexts. Other procedures for combining classifiers consider inputs generated by the contributing classifiers. In another method to combine multiple classifiers, the scores generated by the contributing classifiers are taken as inputs to the combination function. Whichever approach to combination employed, the creation of enhanced classifiers from a set of text classifiers relies on understanding how different classifiers perform in different informational contexts.

## 3. TECHNOLOGIES USED

The main processes involved in this are feature extraction, classification and identification. For this purpose we made use of certain programming platforms.

### 3.1. Java

For performing data pre-processing and feature extraction, Java Development Kit (JDK) and software development kit (SDK) are used. It contains a Java compiler, a full copy of the Java Runtime Environment (JRE), and many other important development tools. MySQL JDBC Driver and edu.mit.jwi_2.1.4 are the two main libraries used for pre-processing data.

### 3.2. Net Beans IDE

Net Beans IDE is used to quickly and easily develop Java application. It provides support for Java Development Kit 7. Here Net Beans Integrated Development Environment runs on the Java SE

Development Kit (JDK) which consists of the Java Runtime Environment plus developer tools for compiling, debugging, and running the applications written in the Java language.

### 3.3. Mat lab

MATLAB helps to perform matrix manipulations, implementation of algorithms, to plot certain functions and data creation of user interfaces etc. It also helps to create interface with other programming languages such as C, C++, Java, and Fortran. Although MATLAB is intended primarily for numerical computing, it consists of an optional toolbox that uses the MuPAD symbolic engine, that allows access to the symbolic computing capabilities. Here MATLAB is used to classify the features extracted. The features extracted by the java programs are written to a text file. Later this text file is accessed by the mat lab program for training the classifiers and to perform author identification.

## 4. SYSTEM ANALYSIS

### 4.1. Data Pre- processing

Data pre- processing is a very important step in authorship attribution. Text documents in their original form are not in suitable form for learning. They must be converted into a suitable input format. It can be converted in to a vector space since most of the learning algorithms use the attribute- value representation. This step is crucial in determining the quality of the next stages, that is, the feature extraction and classification stage. Her data pre- processing involves tokenization and stemming.

#### 4.1.1. Tokenization

Tokenization is the method of splitting a stream of text input into meaningful elements. These meaningful elements are called tokens like symbols, phrases, words so on. The extracted group of tokens act as an input for further processes like parsing and text mining. It is a part of lexical analysis. In languages that use inter-word spaces, this approach is fairly straightforward. Tokenization is particularly difficult for languages such as Chinese which have no word boundaries. But it is easy in the case of the language English.

Usually, the tokenization process occurs at the word level. Yet to define what is meant by a "word" is sometimes difficult to deal with. Often a tokenizer relies on some simple heuristics. This can be made clear with some examples such as:

• All adjacent strings of alphabetic characters are always a part of one token. The same is the case numbers.
• Tokens may be separated by whitespace characters. These may include punctuation characters, a line break or space.
• The resulting list of tokens may or may not include Punctuation and whitespace.

#### 4.1.2. Stemming

Stemming is the process of reducing the inflected words to their root or base form known as stem. The stem may not be same as the morphological root of that word. It is enough that related words map to the same stem, even if that stem is not a convincing root. The programs used for performing stemming are usually referred to as stemming algorithms or stemmers. A stemming algorithm reduces the words "fishing", "fished", "fish", and "fisher" to the root or base form of the word, that is the word "fish".

Here Wordnet Stemmer is used for stemming. This stemmer adds functionality to the simple pattern-based stemmer SimpleStemmer by checking to see if possible stems are actually present in Wordnet. Stems are returned only if any possible stems are found otherwise the word is considered unknown, and the result returned is the same word as that of the SimpleStemmer class. Wordnet dictionary is required to construct a WordnetStemmer.

### 4.2. Feature Extraction

The features and their extraction process are very dependent on the text language. These features can be used to understand the peculiarity of an author' s writing. These features are extracted from the author' s text. Some of the important features extracted here are,

1. Number of periods.
2. Number of commas.
3. Number of question marks.
4. Number of colons.
5. Number of semi- colons.
6. Number of blanks.
7. Number of exclamation marks.
8. Number of dashes.
9. Number of underscores.
10. Number of brackets.
11. Number of quotations.
12. Number of slashes.
13. Number of words.
14. Number of sentences.
15. Number of characters.
16. Ratio of characters in a sentence.
17. Top K word frequency.

### 4.3. Applying Classifiers

After performing the feature extraction process, the extracted features are used to classify the input text data. Mainly two classifiers are used and they are Fuzzy learning classifier and SVM classifier. Then these two classifiers are combined to form a new classifier.

#### 4.3.1. Fuzzy learning classifier

Fuzzy classification is the process of grouping elements into a fuzzy set. Hence, fuzzy classification can be defined as the process of grouping individuals having the same characteristics into a fuzzy set. In this project the texts of different authors are grouped based on the characteristics of each author. In the fuzzy classification technique, there will be a membership function µ that indicates whether an individual is a member of a class. Naturally, a class is a set that is defined by a specific property. Then all objects having that property are the elements of that particular class. The classification process evaluates a given set of objects and checks whether they accomplish the classification property. If it matches then it's a member of the corresponding class.

Here fuzzy classifier is used to classify the given input text and identify the author of that particular text. For this purpose we take numerous texts of each authors and extract the features which are specific to each authors. So each author will have a set of features which are unique to him. So whenever we give a new text or want to identify the author of an unknown text, the features of that text is extracted and checked for matching with any from the poll of author's features. The most likely author is then assigned.

#### 4.3.2. Support Vector Machine classifier
Support Vector Machine is an advanced supervised modelling technique for classifying both linear and nonlinear data. SVM has become a more recent default approach to classification problems since it is well suited to very high-dimensional spaces and extremely big datasets. In machine learning, support vector machines are supervised learning models with associated learning algorithms. They perform actions such as analyze data and recognize patterns. They are used for regression analysis and classification. The simple SVM takes a set of input data. Then for each given input, it predicts, which of two possible classes forms the output which makes it a non-probabilistic binary linear classifier. But here since a group of authors are to be dealt with a multi SVM is used. This SVM is implemented with RBF kernel.

First the support vector machine is trained using the training value sets prepared according to our authors and their texts. The SVM used here is multi SVM. It is because a pool of authors are used. So more than one class is strictly required. After training, the trained machine is used to classify or predict a new input text data. In order, to obtain satisfactory predictive accuracy, various SVM kernel functions are available. The SVM kernel function used here is RBF kernel.

#### 4.3.3. Combined classifier

By combining the output of two classifiers we get a more accurate result than a single classifier. Here two single classifiers were used. They are fuzzy classifier and Support Vector Machine classifier. These two classifiers are then combined to create a new combined classifier. Both the classifiers are executed. The results of both the individual classifiers are compared. This helps to achieve a result with better accuracy compared to individual classifier results.

### 4.4. Author Identification

In the author identification step as the name specifies the author is identified. For a given input text the name of the author of the text will be returned. For this purpose several steps are performed. These steps are the same steps that are performed on the initial stages to differentiate the features of authors from their known texts.

In the identification step, when an unknown or disputed author input text is given, three main steps are performed. Firstly data pre- processing is performed which involves tokenizing and stemming of the input text. Secondly, Feature extraction is performed. Thirdly, after extracting the features classification is done. Any classifier can be chosen to identify the author namely fuzzy, SVM and their combination. But the combined one gives a more accurate result. After that SVM have a better performance than Fuzzy classifier.

## 5. Results and discussions

The experiments used a set of writings of distinct authors. This set comprises many containing many words. The texts are written in English. A set of random paragraphs of each authors were taken and processed.

The set of texts undergoes distinct tokenization and techniques. The distinct tokenization methods were used:

• Words and punctuation
• Words, punctuation and additional stylometric features.

Punctuation improves significantly the identification rate, so it was always included. The additional stylometric features used were:
• Number of words
• Number of sentence
• Number of characters

The text classification for authorship attribution analysis based on Fuzzy and SVM classifiers were performed and their performance based on CPU time is calculated. The following graphs explains the time taken by each classifiers.

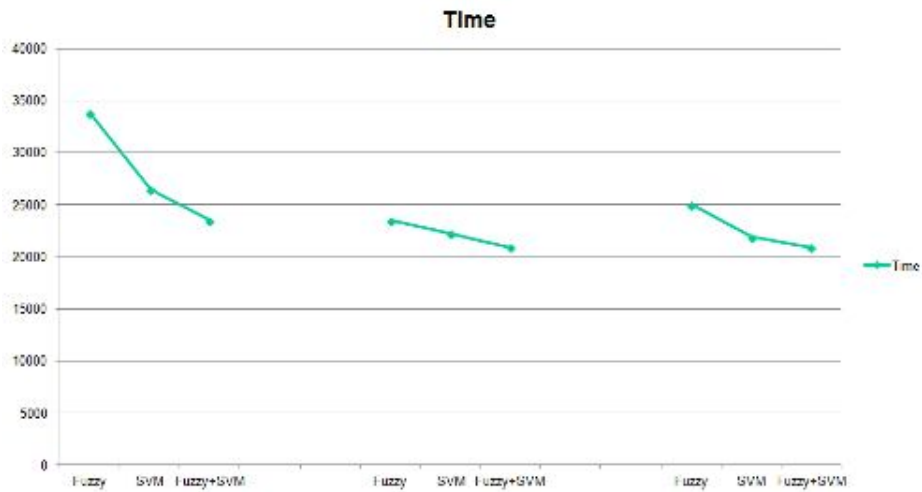

Figure 1. CPU time comparison.

The text classification was performed using different texts of the author from his different writings. 20 different texts of each 10 different authors were taken for this purpose. Thus a total of 200 were used. The accuracy was determined using percentage calculation considering the correctly identified and wrongly identified authors. The following table shows the percentage of different classifiers.

Table 1.  Percentage value of classifiers.

| Sl No | Classifier Type | Percentage value |
|---|---|---|
| 1 | Fuzzy classifier | 58 |
| 2 | SVM classifier | 70 |
| 3 | Fuzzy + SVM classifier | 76 |

The text classification for authorship attribution analysis based on Fuzzy and SVM classifiers has been implemented. Later a combined classifier of these two was implemented. The input given is a randomly selected text of an author. The output obtained is the name of the author. From my work SVM has shown more accuracy than fuzzy technique with an accuracy of 70%. But the combined classifiers gave an accuracy of 76%.

## 6. CONCLUSIONS

The proposed work successfully performs identification of the author of a given input text. In this work different texts of various authors are selected. These texts are then tokenized and stemmed. Then the frequency of each word in the stemmed text is determined and the top k element is chosen. Other features such as number of characters, words, sentences and their ratios are also extracted. The number of different types of punctuations and symbols are also determined. Later these features were analysed to conclude that each author have certain features which are peculiar to himself. Then these features were used to train the fuzzy and SVM classifiers. And the experiments proved that SVM shows more accuracy than fuzzy classifier. Later the combined classifier is found to have more accuracy than the other two individual classifiers.

## ACKNOWLEDGEMENTS

We would like to thank our project guide Mr. M. Sudheep Elayidom, Associate Professor at Division of computer science and Engineering, School of Engineering, Cochin University, for his utmost guidance in our project work.

**Authors**

**M. Sudheep Elayidom**

The author is an Associate Professor in Cochin University of Science and Technology in Kerala, India in the Computer Science and Engineering division, of school of engineering. He received his PhD in computer science from Cochin University, Kerala, India and masters in computer and information science with a first rank from the same university. He got his B.Tech with a first rank from M.G University, Kerala, India. He has published many international papers in the domain of data mining and is active in the research area guiding research scholars in the fields of big data, cloud data bases and so on.

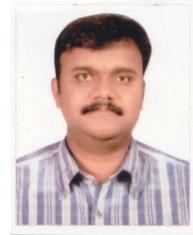

**Chinchu Jose**

The author did B.Tech in Computer Science and Engineering from M.G University, Kerala, India. Now pursuing M.Tech in Computer science and Engineering from M. G university.

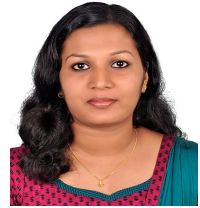

**Anitta Puthussery**

The author did B.E in Computer Science and Engineering from Anna University, Tamilnadu, India. Now pursuing M.Tech in Computer science and Engineering from M. G university.

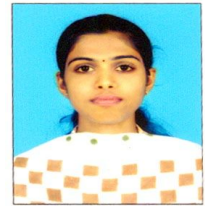

**Neenu K Sasi**

The author did B.Tech in Computer Science and Engineering from M.G University, Kerala, India. Now pursuing M.Tech in Computer science and Engineering from M. G university.

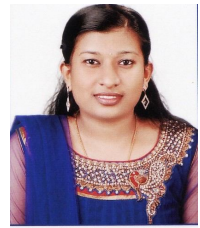